\documentclass[twocolumn,showpacs,prb]{revtex4}
\usepackage{epsfig}
\usepackage{graphicx}
\usepackage{dcolumn}
\usepackage{bm}

\begin{document}

\title{Microscopic origin of the linear temperature increase of the magnetic susceptibility of BaFe$_{2}$As$_{2}$}
\author{S.L.~Skornyakov$^{1,2}$, V.I.~Anisimov$^{1,2}$, and D.~Vollhardt$^{3}$}
\affiliation{$^{1}$Institute of Metal Physics, Russian Academy of Sciences, 620041
Yekaterinburg GSP-170, Russia \\
$^{2}$Ural Federal University, 620002 Yekaterinburg, Russia\\
$^{3}$ Theoretical Physics III, Center for Electronic
Correlations and Magnetism, Institute of Physics, University of
Augsburg, D-86135 Augsburg, Germany}
\date{\today }

\begin{abstract}
Employing a combination of \emph{ab initio} band structure theory and dynamical mean-field theory
we explain the experimentally observed linear temperature increase of the magnetic susceptibility
of the iron pnictide material  BaFe$_{2}$As$_{2}$. The microscopic origin of this anomalous behaviour
is traced to a sharp peak in the spectral function located approximately 100~meV below the Fermi
level. This peak is due to the weak dispersion of two-dimensional bands associated with the layered
crystal structure of pnictides.
\end{abstract}

\maketitle

\section{INTRODUCTION}
Since the discovery of high-temperature superconductivity in the iron pnictides \cite{Fepnic}
the unusual electronic properties of this class of materials has attracted considerable attention
\cite{singhsdw,rotter,Wang,klingeler,korshunov,zhang,Aichhorn,BaFe2As2,LaFePO10,arpes1,arpes2,skornyakov}.
The interest was stimulated by the fact that the new superconductors share several similarities
with the well-studied, but still not well understood, high-$T_c$ cuprates. Firstly, both classes of
superconductors crystallize into a layered structure. Secondly, in most cases the parent compounds
of the pnictides are not superconducting, and superconductivity emerges only under doping or
pressure and is associated with the suppression of antiferromagnetic (AFM) order. However, in
contrast to the cuprates whose parent compounds are Mott insulators, the parent compounds of the
pnictides are multiband metals. The nature of the magnetic ground state of the parent compounds
is also different: in the cuprates it corresponds to a N\'eel-type order of a Mott-Hubbard insulator,
while in the pnictides magnetism is associated with a nesting-induced spin density
wave\cite{singhsdw, rotter, delacruz} (SDW).

The magnetic properties of pnictide materials show anomalous behavior even in the paramagnetic state.
An unusual linear temperature increase of the uniform magnetic susceptibility was reported in the
parent compound BaFe$_{2}$As$_{2}$\cite{Wang} as well as in stoichiometric and fluorine-doped
LaFeAsO\cite{klingeler}. It is now well established that the linear increase of the magnetic
susceptibility with temperature is a general property of all pnictide superconductors for temperatures
above the SDW transition. Nevertheless, no consensus has been reached so far about the origin of this
phenomenon. To date several mechanisms were proposed to explain the observed $T$-dependence in the
pnictides. Wang {\it et al.}\cite{Wang} and Zhang {\it et al.}\cite{zhang} suggested that the linear-$T$
behavior is a consequence of strong antiferromagnetic fluctuations present above the SDW transition
temperature. Korshunov {\it et al.} \cite{korshunov} argued that short-range antiferromagnetic
fluctuations are the source of a linear-$T$ term in the susceptibility of a two-dimensional Fermi
liquid which allowed them to obtain good agreement with experimental data.

A very important issue concerning the spectral and magnetic properties of the pnictides is the role of
Coulomb correlations. It is now generally accepted that electronic correlations in the pnictides
are not as strong as in the cuprates and should be classified as moderate\cite{Aichhorn, BaFe2As2}.
It was shown\cite{BaFe2As2,LaFePO10} that the spectral properties of the pnictides can be reproduced
by first-principles techniques only if local dynamical Coulomb correlations are taken into account.
This can be achieved by employing the LDA+DMFT method\cite{LDApDMFT}. This computational scheme combines
electronic band structure calculations in the local density approximation (LDA) with many-body physics
incorporated in the dynamical mean-field theory (DMFT)\cite{DMFT89}.

In our earlier study\cite{skornyakov} we proposed an explanation of the temperature increase of the
magnetic susceptibility of LaFeAsO which was based on a first-principles analysis of the low-energy
spectral properties caused by local dynamical Coulomb correlations, without taking into account
interatomic magnetic fluctuations. In the present paper we employ the LDA+DMFT scheme to demonstrate
that the proposed mechanism can be applied also to understand the origin of the linear temperature
dependence of the uniform magnetic susceptibility in stoichiometric BaFe$_{2}$As$_{2}$.

\section{COMPUTATIONAL METHOD}
The LDA+DMFT computational scheme implemented in the present work proceeds in four steps: (i) the
construction of an effective tight-binding Hamiltonian $\hat H^{\rm WF}({\bf k})$ from a converged
LDA solution by projecting\cite{proj} onto Wannier functions, (ii) the addition of the local Coulomb
interaction $\hat H^{\rm Coul}$, (iii) a double-counting correction which takes into account the local
interactions already decribed by the LDA, and (iv) the self-consistent solution of the DMFT equations
on the Matsubara contour with continuous-time quantum Monte-Carlo\cite{CTQMC1} (CT-QMC) as impurity
solver. The Hamiltonian to be solved by DMFT is given by
\begin{eqnarray}
\hat H^{\rm DMFT}({\bf k})=\hat H^{\rm WF}({\bf k}) + \hat H^{\rm Coul} + \hat H^{\rm DC}.
\label{H1}
\end{eqnarray}
Exact investigations of the magnetic response of this model must employ a rotationally invariant form of
the interaction term $\hat H^{\rm Coul}$. However, up to now there do not exist effective algorithms for the solution of a five-orbital Hubbard model with the full Coulomb interaction within CT-QMC. Furthermore, our investigation of the magnetic properties of
BaFe$_{2}$As$_{2}$ requires a separate, and extremely time-consuming, \emph{self-consistent}
DMFT calculation for each point of the susceptibility curve. To make computations feasible, we therefore include only the density-density contributions to the full interaction
\begin{equation}
\hat H^{\rm Coul} \equiv \hat H^{\rm U}=\frac{1}{2}\sum_{i,\alpha,\alpha^{\prime},\sigma,\sigma^{\prime}}
U^{\sigma\sigma^{\prime}}_{\alpha\alpha^{\prime}}
\hat n_{i\alpha\sigma}^{d}
\hat n_{i\alpha^{\prime}\sigma^{\prime}}^{d}.
\end{equation}
Here $U^{\sigma\sigma^{\prime}}_{\alpha\alpha^{\prime}}$ is the Coulomb interaction matrix,
$\hat n_{i\alpha\sigma}^{d}$ is the occupation number operator for $d$ electrons in the orbital
$\alpha$ or $\alpha^{\prime}$, with spin $\sigma$ or $\sigma^{\prime}$, on the $i$th site.
This approximation neglects spin flip and pair hopping processes. Nevertheless, as will be shown below, it is able to provide correct results for the spectral and magnetic properties of BaFe$_{2}$As$_{2}$.
The double-counting term is $\hat H^{\rm DC}=-{\bar U}(n_{\rm DMFT}-\frac{1}{2})\hat I$, where
$n_{\rm DMFT}$ is the total, self-consistently determined number of $d$ electrons obtained
within LDA+DMFT, and $\bar U$ is the average Coulomb parameter for the $d$ shell.

We construct Wannier functions in the energy window including Fe-$d$ and As-$p$
states. Hence, by construction energy bands of the $H^{\rm WF}$ Hamiltonian
exactly reproduce 16 Fe-$d$ and As-$p$ bands (two As and Fe atoms in the formula unit, one
formula unit in the unit cell) obtained in LDA calculations, and the $p-d$ hybridization is explicitly
taken into account.

The interaction matrix $U^{\sigma\sigma^{\prime}}_{\alpha\alpha^{\prime}}$ is parametrized by the effective
on-site Coulomb parameter $U$ and intra-orbital exchange parameter $J$ according to the procedure
described in Ref.\cite{UJCalculation}. In the present calculation we use $U$=3.5~eV and $J$=0.85~eV
obtained with the constrained DFT procedure\cite{ConstrainedDFT1,ConstrainedDFT2,ConstrainedDFT3}.

The  orbitally resolved spectral functions  of the interacting system are then computed as
\begin{equation}
A_{\alpha}(\omega )=-\frac{1}{\pi }\mathrm{Im}\sum_{{\mathbf k}}[(\omega +\mu )
\hat{I}-\hat{H}^{\mathrm{WF}}(\mathbf{k})-\hat{H}^{\mathrm{DC}}-\hat{\Sigma}(\omega )]_{\alpha\alpha}^{-1}, \nonumber
\label{SFunction}
\end{equation}
where the subscript $\alpha$ refers to an orbital, $\mu$ is the self-consistent chemical potential, and
$\Sigma (\omega)$ is the self-energy on the real axis obtained by analytic continuation
using the Pad\'e approximant\cite{Pade1} technique; details are described in Ref.\cite{LaFePO10}.

The uniform magnetic susceptibility is calculated as the response of the system to a weak external
magnetic field,
\begin{equation}
\chi (T)=\frac{\partial M(T)}{\partial E_{h}}=\frac{\partial \lbrack
n_{\uparrow }(T)-n_{\downarrow }(T)]}{\partial E_{h}},  \nonumber
\label{chidefinition}
\end{equation}
where $M(T)$ is the field-induced magnetization, $n_{\sigma}(T)$ is the number of electrons with
spin $\sigma$, and $E_{h}$ is the energy correction corresponding to the applied field. Since the
field is finite the calculations are performed in three steps: First, we check that the polarization
is zero in the absence of the field, then we check that $M(T)$ is a linear function of $E_{h}$, and
finally we evaluate the derivative in Eq.~(\ref{chidefinition}) as a ratio of $M(T)$ and $E_{h}$.

\section{RESULTS}
\subsection{Temperature dependence of the uniform magnetic susceptibility}
\begin{figure}[t]
\centering \vspace{0.0mm}
\includegraphics[width=0.90\linewidth,angle=0]{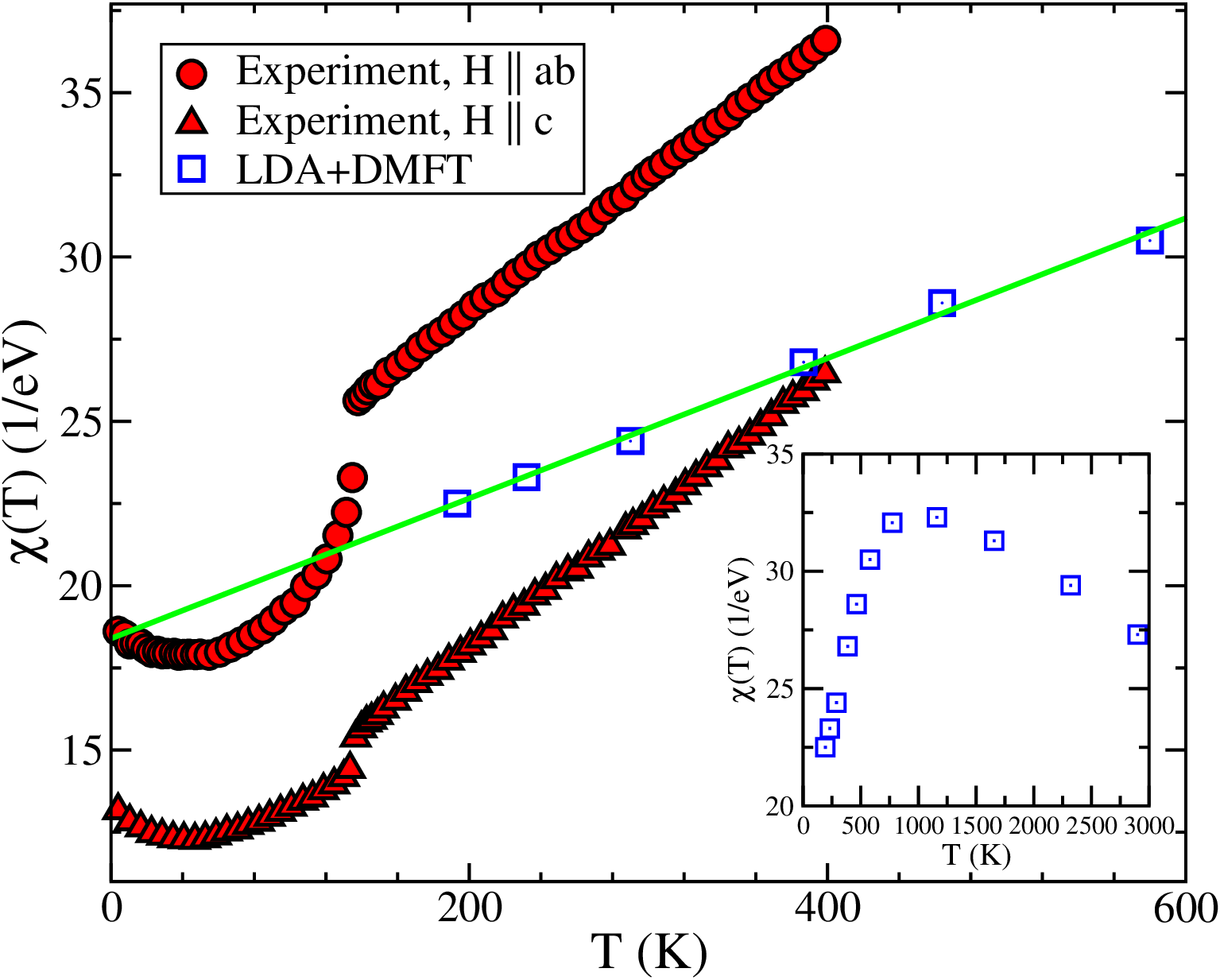}
\vspace{0.0mm}
\caption{(Color online) Uniform magnetic susceptibility $\chi(T)$ of BaFe$_{2}$As$_{2}$
calculated within LDA+DMFT~(squares) in comparison with experimental
data of Wang {\it et al}.\cite{Wang} (circles and triangles). The line is a least-square fit to
the last six points of the computed data. The inset shows the theoretical curve for the full
temperature interval.}
\label{DMFTsuscvsExp}
\end{figure}

In Fig.~\ref{DMFTsuscvsExp} the uniform magnetic susceptibility $\chi(T)$ computed within LDA+DMFT
is compared with the experimental data of Wang {\it et al.}\cite{Wang}. In the temperature range
from 200~K to 600~K the temperature dependence of the calculated $\chi(T)$ is found to be almost
perfectly linear. However, the slope is by a factor of 1.7 smaller than in the experiment.
The origin of this quantitative discrepancy is not clear at the moment and will be the subject of
future investigations. To emphasize the linearity we plot a least-square fit to the last six points
of the computed data. We note that the observed linear behavior is, in fact, due to an extended linear
region around the turning point of $\chi(T)$ at $\approx$350~K. The obtained $\chi(T)$ has a maximum
at about 1000~K and decreases for higher temperatures.

\begin{figure}[h]
\centering \vspace{0.0mm}
\includegraphics[width=0.90\linewidth,angle=0]{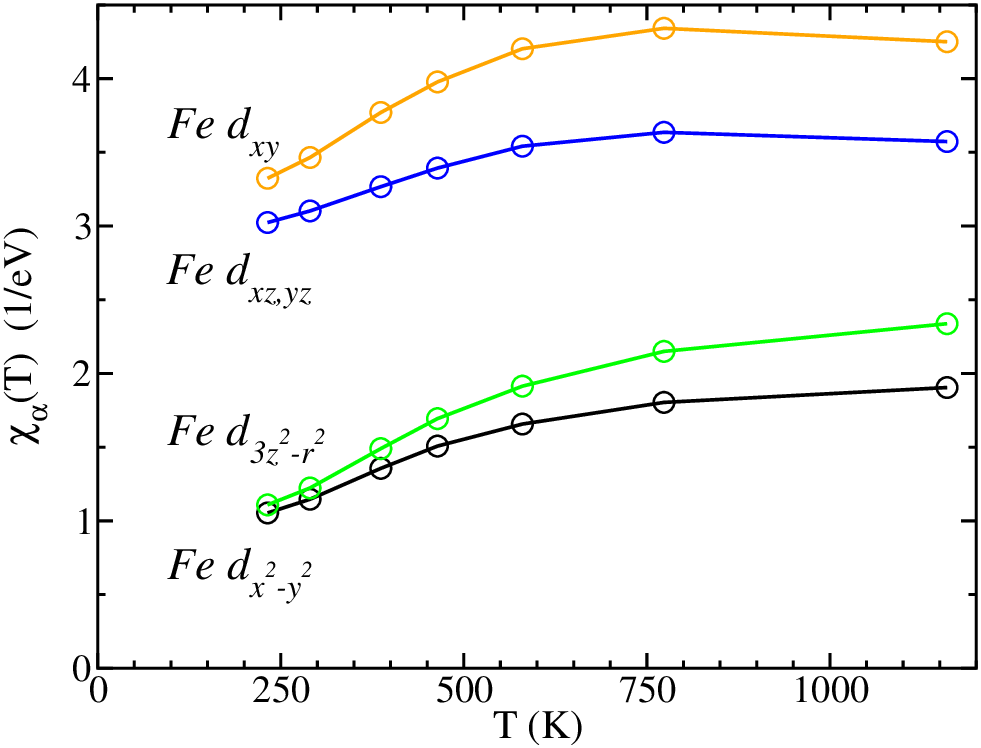}
\vspace{0.0mm}
\caption{(Color online) LDA+DMFT results for the orbitally resolved  Fe~3$d$ susceptibilities
$\chi_{\alpha}(T)$ with $\alpha=xy$, $yz$, $xz$, $3z^{2}-r^{2}$, $x^{2}-y^{2}$ of BaFe$_{2}$As$_{2}$ vs.
temperature obtained from the derivative of the magnetization.}
\label{SuscepDirectResolved}
\end{figure}

More detailed information about the magnetic properties of BaFe$_{2}$As$_{2}$ can be obtained from
an analysis of the orbitally resolved contributions $\chi_{\alpha}(T)$ ($\alpha=xy$, $yz$, $xz$, $3z^{2}-r^{2}$,
$x^{2}-y^{2}$) to the total magnetic susceptibility. In Fig.~\ref{SuscepDirectResolved} we show the
temperature dependence of the Fe~3$d$ susceptibilities of BaFe$_{2}$As$_{2}$ obtained in LDA+DMFT.
All contributions have approximately equal slope in the temperature interval from 200~K to 500~K.
The $d_{xy}$ orbital of Fe provides the largest contribution to the total susceptibility.

\subsection{Connection between magnetic and spectral properties}
\begin{figure}[b]
\centering \vspace{0.0mm}
\includegraphics[width=0.90\linewidth,angle=0]{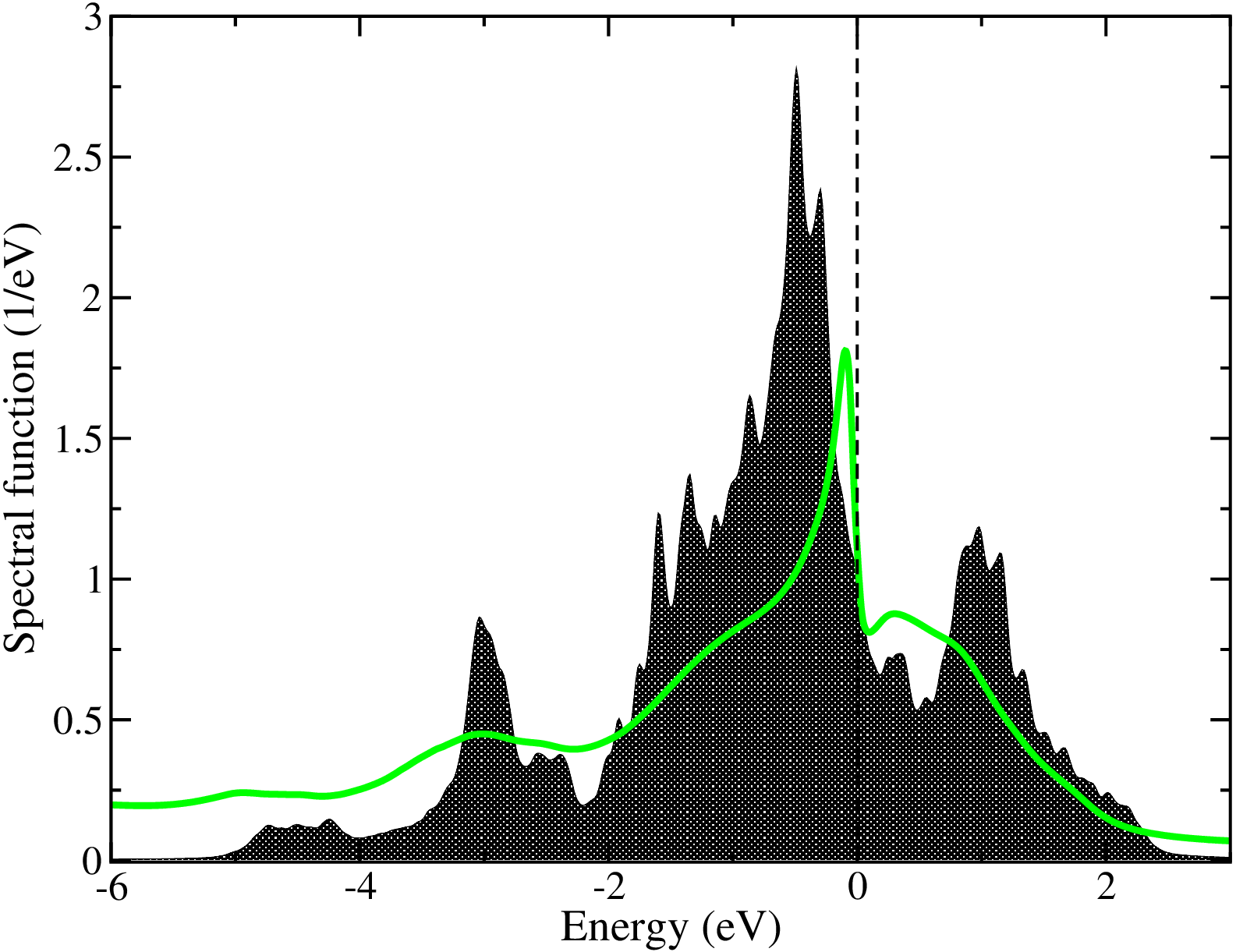}
\caption{(Color online) Comparison of the total Fe 3$d$ spectral function of BaFe$_{2}$As$_{2}$
as obtained from LDA+DMFT (solid line) and LDA (shaded area), respectively.}
\label{AwDMFTvsLDA}
\end{figure}

In our previous paper\cite{skornyakov} we proposed a scenario according to which the anomalous
$T$-behavior of $\chi(T)$ in the pnictides is connected with the presence of a sharp peak in the
spectral function below the Fermi energy. In Fig.~\ref{AwDMFTvsLDA} the total Fe 3$d$ spectral
function $A(\omega)$ computed within LDA+DMFT is shown in comparison with the LDA result. It
demonstrates that dynamical correlation effects strongly renormalize the spectrum in the vicinity
of the Fermi energy. In particular, the electronic correlations are seen to lead to a narrow peak
below the Fermi level while the remaining part of the spectrum is only weakly affected by the
correlations. The peak in the energy window from -4~eV to -2~eV should not be mistaken for a lower
Hubbard band since a similar peak is already present in the LDA result. The Fe 3$d$ spectral weight in
this energy area is a consequence of the hybridization with As 4$p$ states.

\begin{figure}[b]
\centering \vspace{0.0mm}
\includegraphics[width=0.90\linewidth,angle=0]{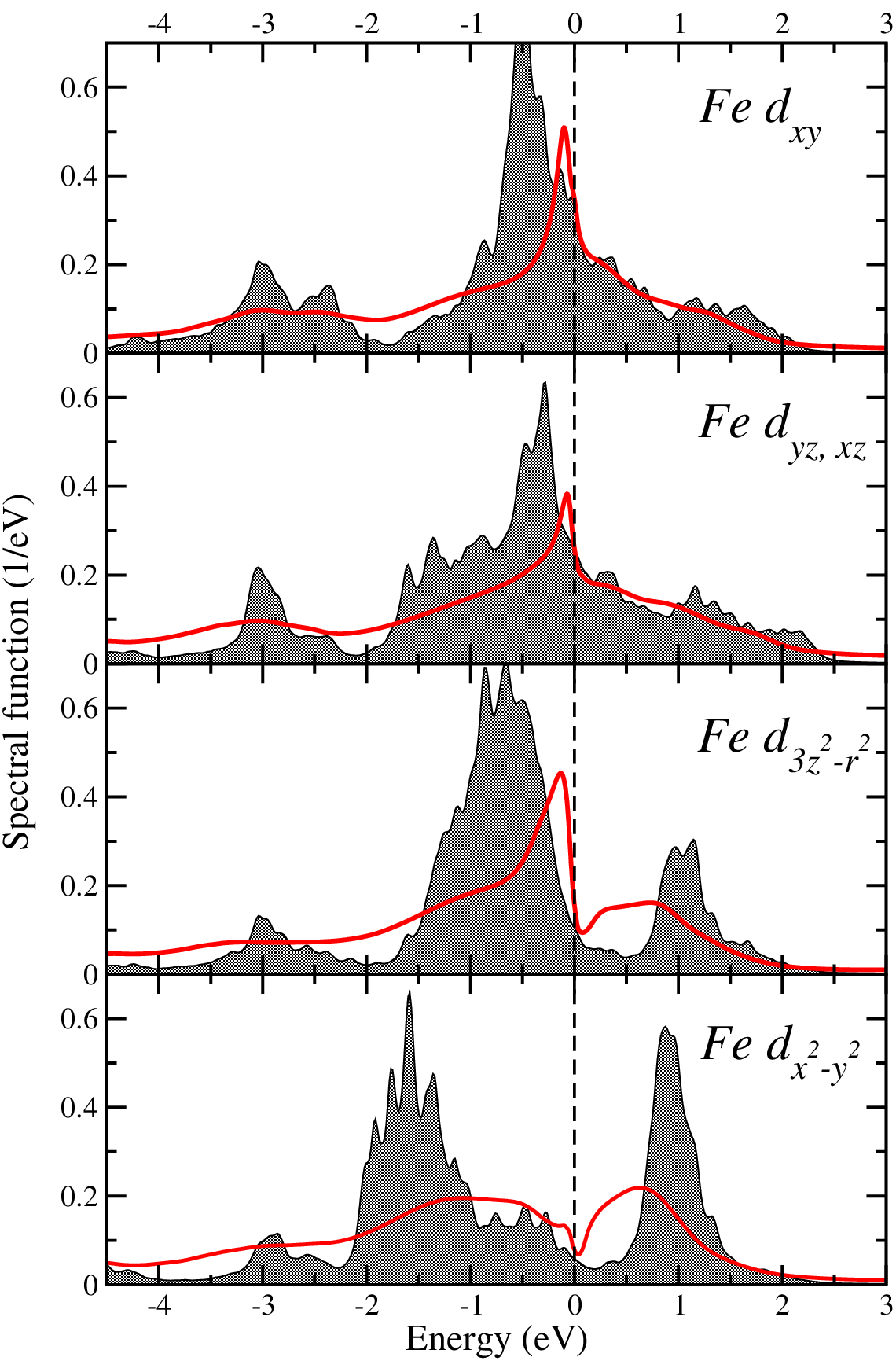}
\caption{(Color online) Orbitally resolved Fe 3{\it d} spectral functions of
BaFe$_{2}$As$_{2}$ obtained within LDA+DMFT (solid lines) in comparison with
LDA results (filled areas). The Fermi energy is set to 0~eV.}
\label{DMFTvsLDA}
\end{figure}

A comparison of the orbitally resolved spectral functions computed within LDA and LDA+DMFT,
respectively, is shown in Fig.~\ref{DMFTvsLDA}. The LDA+DMFT results are in good agreement with
previously reported theoretical and experimental spectra \cite{BaFe2As2}. Except for the $d_{x^{2}-y^{2}}$
orbital the spectral functions obtained by LDA+DMFT all show a sharp peak below the Fermi energy.
These peaks originate from local correlation effects as pointed out above.

A quantitative measure of the correlation strength is the quasiparticle renormalization factor $Z$.
In the single-orbital case it can be obtained from the real-axis self-energy $\Sigma(\omega)$ which
is related to the effective mass enhancement $m^{*}/m$ by $Z^{-1}=1-\partial Re(\Sigma(\omega))/\partial\omega=m^{*}/m$.
For a multi-orbital problem the self-energy is a matrix. Therefore different orbitals have different
$m^{*}/m$. The computed values of the mass enhancement range from 2.5 to 3.7 and agree well with previous
estimates of $m^{*}/m$ for pnictides obtained from the renormalization of the LDA band structure
\cite{Aichhorn,BaFe2As2,LaFePO10,arpes1,arpes2}.

\begin{figure}[b]
\centering \vspace{0.0mm}
\includegraphics[width=0.90\linewidth,angle=0]{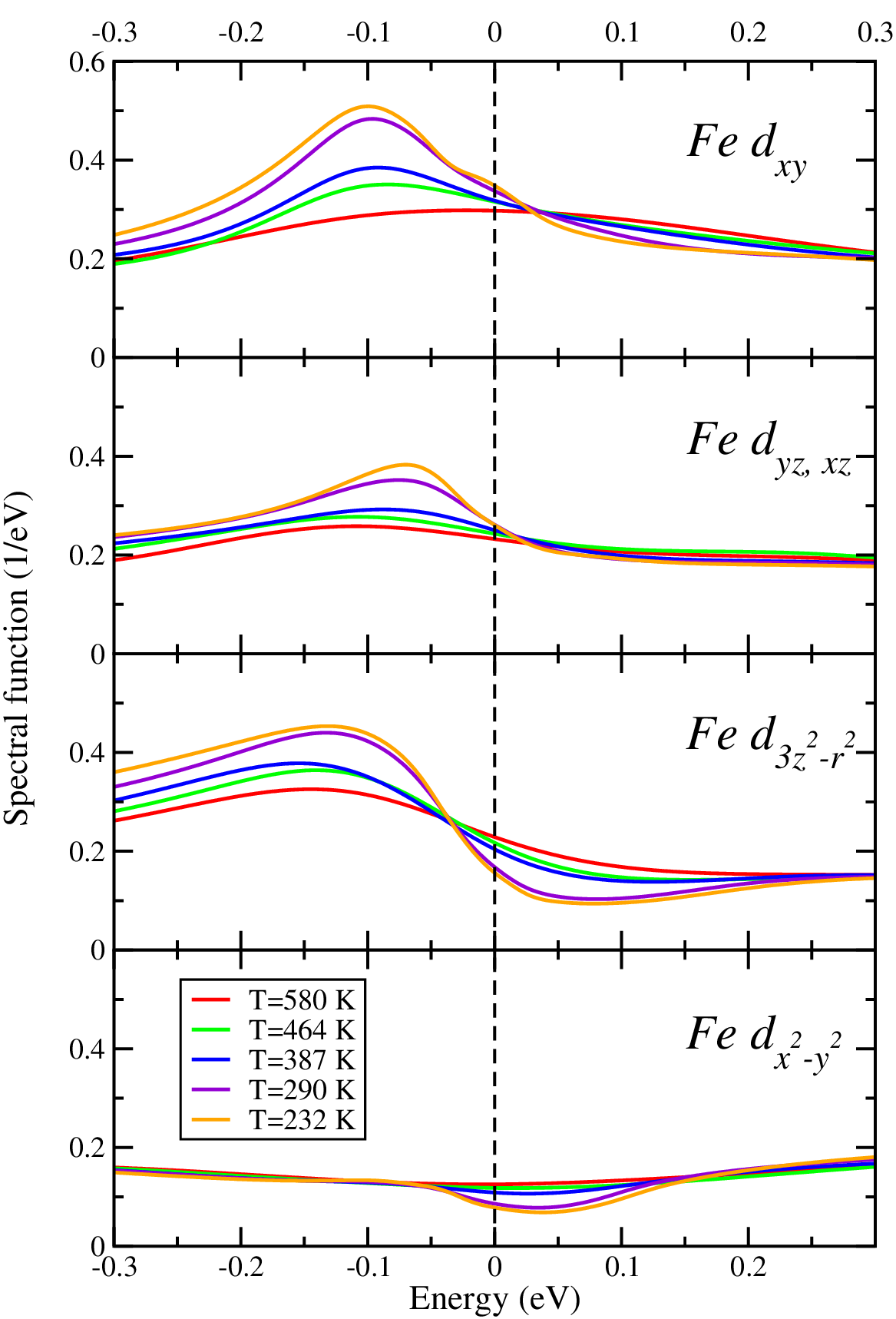}
\caption{(Color online) Fe~3{\it d} spectral functions of BaFe$_{2}$As$_{2}$ computed within LDA+DMFT
in the temperature range 232~K - 580~K. The Fermi energy corresponds to 0~eV.}
\label{TDMFT}
\end{figure}

To identify possible reasons for the anomalous behavior of the magnetic properties it is instructive
to compute the temperature evolution of the orbitally resolved spectral functions. The spectral
curves computed for temperatures ranging from 232~K to 580~K are shown in Fig.~\ref{TDMFT}.
All spectral functions, except for the $d_{x^2-y^2}$ orbital, show a temperature sensitive peak which
is located approximately 100~meV below the Fermi energy. These peaks increase in amplitude and become
narrow with decreasing $T$. We note that the magnitudes of the orbital contribution $\chi_{\alpha}(T)$
are proportional to the density of states of the corresponding orbitals at the Fermi energy.

The simplest way to establish a possible connection between the temperature evolution of the magnetic
susceptibility and the excitation spectrum in the multi-orbital case is to estimate the susceptibility
(per spin) using the bubble diagram obtained by convoluting the DMFT Green's functions, Eq.~(\ref{chi0m}),
\begin{equation}
{\chi}^{0}_{\alpha}(T)=\frac{1}{\beta }\sum_{\mathbf{k,}i\omega,\alpha^{\prime}}
\hat{G}_{\alpha\alpha^{\prime}}(\mathbf{k},i\omega )\hat{G}_{\alpha^{\prime}\alpha}(\mathbf{k},i\omega ),
\label{chi0m}
\end{equation}
where
$\hat{G}_{\alpha\alpha^\prime}(\mathbf{k},i\omega )=
[(i\omega +\mu )\hat{I}-\hat{H}^{\mathrm{DMFT}}(\mathbf{k})-\hat{\Sigma}(i\omega )]_{\alpha\alpha^\prime}$.
This expression describes the spin susceptibility in the absence of vertex corrections, i.e., gives
an estimate of the magnetic response due to single-particle excitations which are characterized by
the interacting spectral function. An explicit connection between the magnetic response, Eq.~(\ref{chi0m}),
with the excitation spectrum can be made in the one-orbital case when off-diagonal elements of $\hat{G}$
are absent:
\begin{equation}
\chi_{1}(T)=-\frac{1}{4\pi^2}\int d\omega_{1}d\omega_{2}
\frac{f_{\mathrm{F}}(\omega_{1})-f_{\mathrm{F}}(\omega_{2})}{\omega_{1}-\omega_{2}}
A(\omega_{1})A(\omega_{2}).
\label{chiviados}
\end{equation}
Here $A(\omega)$ is the spectral function of the interacting system(per spin), and the temperature
enters via the Fermi function, $f_{\mathrm{F}}(\omega)$. In real compounds multi-orbital physics, including
hybridization effects, is important, in which case the full matrix Green functions must be used in Eq.~(\ref{chi0m}).

\begin{figure}[h]
\centering \vspace{0.0mm}
\includegraphics[width=0.90\linewidth,angle=0]{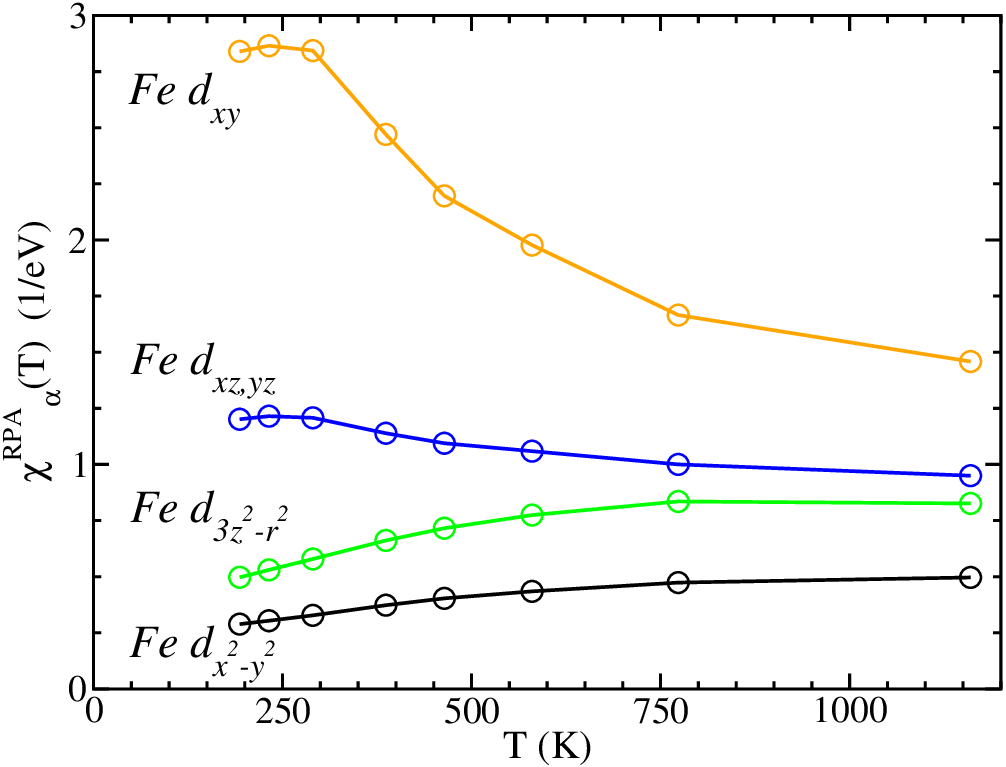}
\vspace{0.0mm}
\caption{(Color online) LDA+DMFT results for the orbitally resolved  Fe~3$d$ susceptibilities
$\chi^{\mathrm{RPA}}_{\alpha}(T)$ with $\alpha=xy$, $yz$, $xz$, $3z^{2}-r^{2}$, $x^{2}-y^{2}$ of BaFe$_2$As$_2$
vs. temperature, obtained from the RPA expression for the susceptibility in Eq.~(\ref{chirpa}).}
\label{SuscepConvResolved}
\end{figure}

The influence of interaction effects may be estimated by calculating the magnetic susceptibility in the
random-phase approximation (RPA). If the orbital dependence of the Coulomb interaction between $d$ electrons
is neglected, the orbital contributions to the total uniform $d$ magnetic susceptibility within RPA are
given by
\begin{equation}
\chi^{\mathrm{RPA}}_{\alpha}(T)=\frac{2\chi^{0}_{\alpha}(T)}{1-{\bar U}\chi^{0}_{\alpha}(T)},
\label{chirpa}
\end{equation}
where the factor 2 is due to the spin degeneracy. The temperature dependence of the orbitally resolved
susceptibilities computed according to Eq.~(\ref{chirpa}), is shown in Fig.~\ref{SuscepConvResolved}.
For the $d_{3z^2-r^2}$ and $d_{x^2-y^2}$ orbitals the results are in good qualitative agreement with the
full LDA+DMFT solution. By contrast, the temperature dependence of the susceptibilities corresponding to
the $d_{xy}$, $d_{yz}$, and $d_{xz}$ states is not reproduced by Eq.~($\ref{chirpa}$); apparently vertex
corrections are important in this case.

The above results suggest the following interpretation, whose correctness will be demonstrated later:
The electronic states forming the sharp peak in the spectral function below the Fermi energy lead to
thermal excitations which contribute to the susceptibility. When the energy $k_{\mathrm B}T$ is larger
than the distance between the peak and the Fermi level, the number of states which can be excited is
reduced and the susceptibility starts to decrease. A more complex mechanism is responsible for the
increase of the $d_{x^2-y^2}$ susceptibility where the corresponding spectral function does not show a
peak below the Fermi energy.

Namely, as in the case of LaFeAsO\cite{skornyakov} the off-diagonal contributions  to the susceptibility,
$\chi^{0}_{\alpha\alpha^{\prime}}=1/\beta \sum_{\mathbf{k,}i\omega}\hat{G}_{\alpha\alpha^{\prime}}(\mathbf{k},i\omega )\hat{G}_{\alpha^{\prime}\alpha}(\mathbf{k},i\omega )$,
and in particular  $\chi^{0}_{x^2-y^2,3z^2-r^2}(T)$, are responsible for the increase of $\chi_{x^2-y^2}^{0}(T)$.
Thus the temperature increase of $\chi^{\mathrm{RPA}}_{x^2-y^2}(T)$ is caused by the magnetic response
of the other orbitals (for details see the supplementary material of Ref.\cite{skornyakov}). A detailed
analysis of the proposed mechanism of the increase of $\chi(T)$ using a simplified model will be
presented in the following section.

\subsection{Model analysis}
\begin{figure}[t]
\centering \vspace{0.0mm}
\includegraphics[width=0.90\linewidth,angle=0]{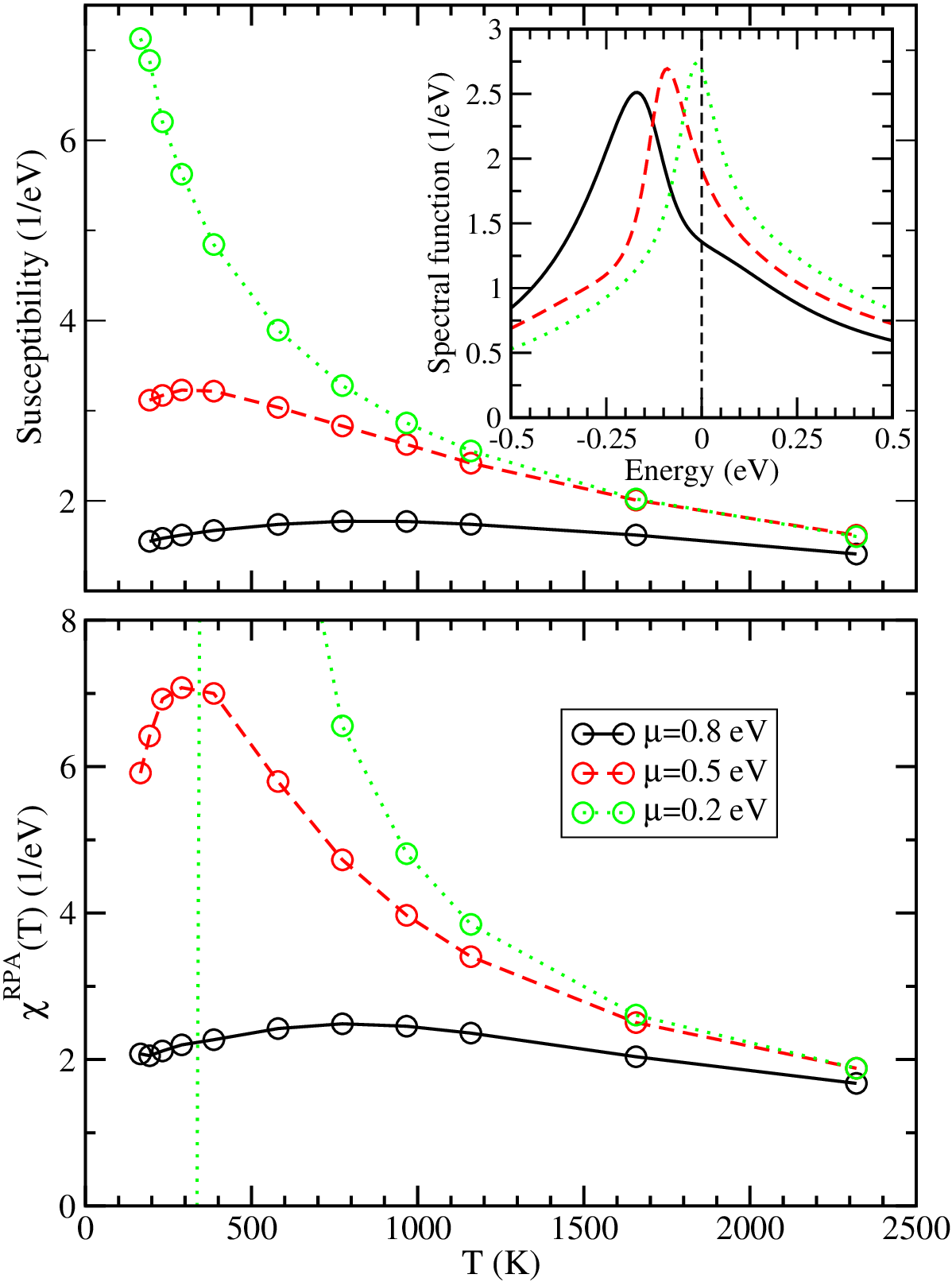}
\caption{(Color online) LDA+DMFT results for the uniform magnetic susceptibility of a one-band model with
the non-interacting Fe $d_{xy}$ DOS of BaFe$_{2}$As$_{2}$ as a function of temperature. Upper panel: Results
computed according to Eq.~(\ref{chidefinition}); the inset
shows the spectral function of the interacting system. Lower panel: RPA results for the uniform magnetic
susceptibility using Eq.~(\ref{chirpa}); the divergence of the susceptibility for $\mu$=0.2~eV at about 300~K
is an artifact of the RPA.}
\label{SuscepDMFTxy}
\end{figure}

To further clarify the relation between the shape of the spectral function and the anomalous temperature
behavior of the magnetic properties we will now perform a model calculation, where multi-orbital effects
are neglected. As a first step we compute the DMFT spin susceptibility for a single-band model, constructed
in such a way that the non-interacting system has exactly the same spectral function as the one for the
$d_{xy}$ orbital of the tight-binding Hamiltonian $H_{\mathrm{WF}}$ in Eq.~(\ref{H1}). In the upper panel
of Fig.~\ref{SuscepDMFTxy} the temperature dependence of the spin susceptibility computed for the model
within DMFT is shown for several values of the chemical potential. The corresponding spectral functions of
the interacting system are presented in the inset of Fig.~\ref{SuscepDMFTxy}.
Depending on the peak position there are two characteristically different temperature dependencies of the
susceptibility: (i) an increase in the low temperature region with a maximum at an intermediate temperature
followed by a decrease at higher $T$ ($\mu$=0.5~eV, 0.8~eV), and (ii) a monotonic decrease with temperature
($\mu$=0.2~eV). The former regime is obtained when the peak is substantially below the Fermi energy, the
latter regime corresponds to the case when the peak is very close to, or right at, the Fermi energy. The
result of the estimate of the susceptibility according to Eq.~(\ref{chirpa}) is presented in the lower panel
of Fig.~\ref{SuscepDMFTxy}. It is seen to  reproduce all features of the curves computed by
Eq.~(\ref{chidefinition}).

This confirms that the magnetic response of the system is indeed governed by its spectral properties. Thus,
the driving force of the non-monotonic behavior of the susceptibility is the peak below the Fermi energy.
In particular, the distance between the peak and the Fermi level is important. In spite of the fact that the peak in the spectral function exists at the level of LDA, its position is too far from the Fermi energy to cause a serious increase of the susceptibility. The correlations shift the peak towards the Fermi level which
results in more pronounced temperature dependence of $\chi(T)$.

\begin{figure}[t]
\centering \vspace{0.0mm}
\includegraphics[width=0.999\linewidth,angle=0]{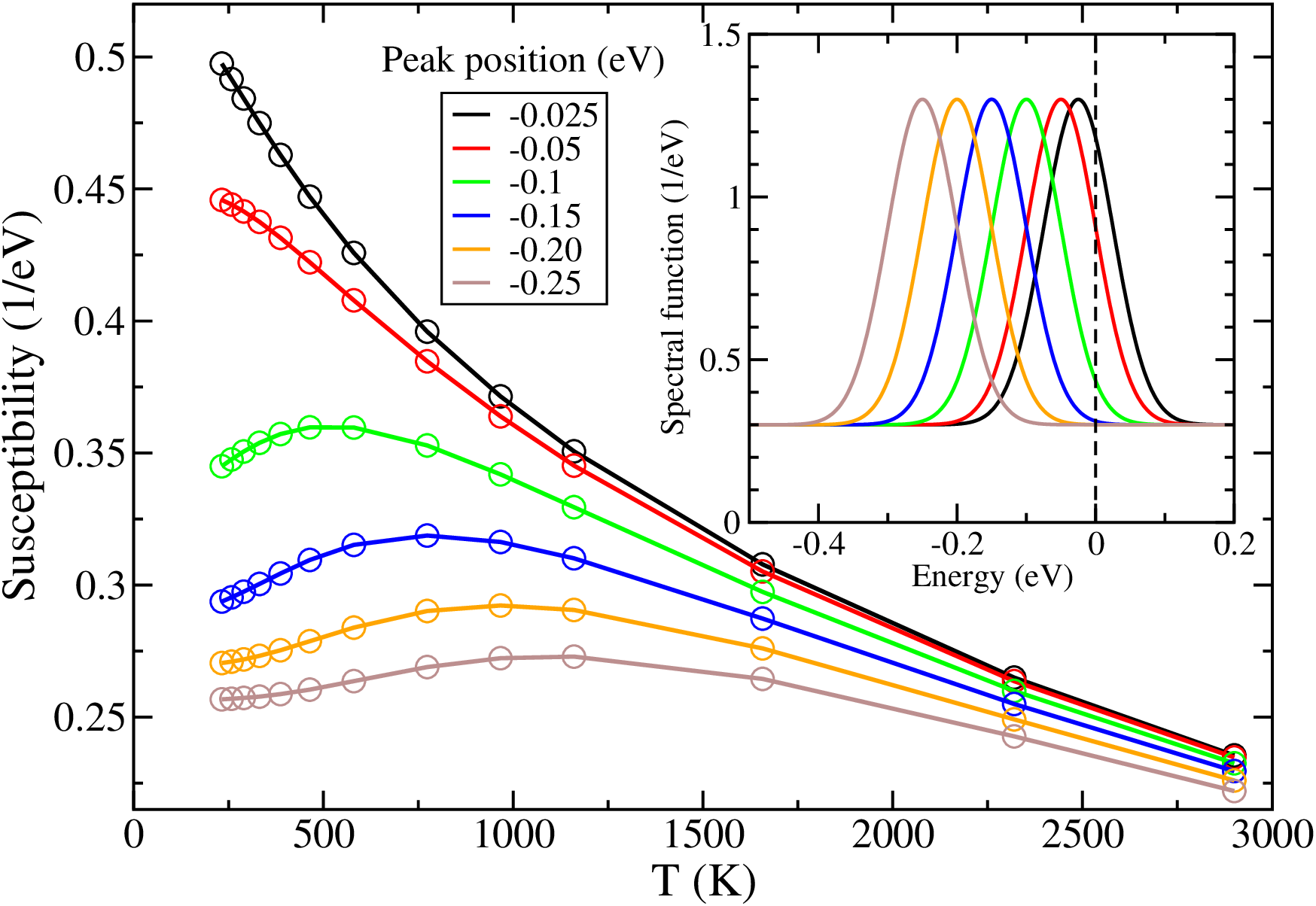}
\caption{(Color online) Spin susceptibility $\chi_{1}(T)$ of a model defined by the density of states
(\ref{DOSGauss}) calculated from Eq.~(\ref{chiviados}) for different values of the peak position. The inset
shows the density of states plotted relative to the Fermi energy (zero energy).}
\label{SuscepGauss}
\end{figure}
To relate the shape of the spectral function with the two temperature regimes of the magnetic susceptibility
even more explicitly, we now compute the temperature behavior of $\chi^{0}(T)$ for a system with a density
of states given by a Gauss function with offset from the energy axis,
\begin{equation}
A(\omega)=A+B\exp[(\frac{\omega-C}{2\sigma})^2].
\label{DOSGauss}
\end{equation}
The parameters in Eq.~(\ref{DOSGauss}) were adjusted such that the maximum and the width of the function
are close to the ones obtained with LDA+DMFT for the material specific Hamiltonian for BaFe$_{2}$As$_{2}$.
Results for $\chi_{1}(T)$ computed according to Eq.~(\ref{chiviados}) are shown in Fig.~\ref{SuscepGauss}.
The temperature behavior of $\chi_{1}(T)$ and its evolution upon changes of the peak position in the spectral
function are seen to qualitatively reproduce all features obtained in DMFT for the one-band model. This
can be viewed as a direct indication that the peculiarities observed in the anomalous behavior of the spin
susceptibility in DMFT originate from the shape of the spectral function in the vicinity of the Fermi energy.

\begin{figure}[t]
\centering \vspace{0.0mm}
\includegraphics[width=0.90\linewidth,angle=0]{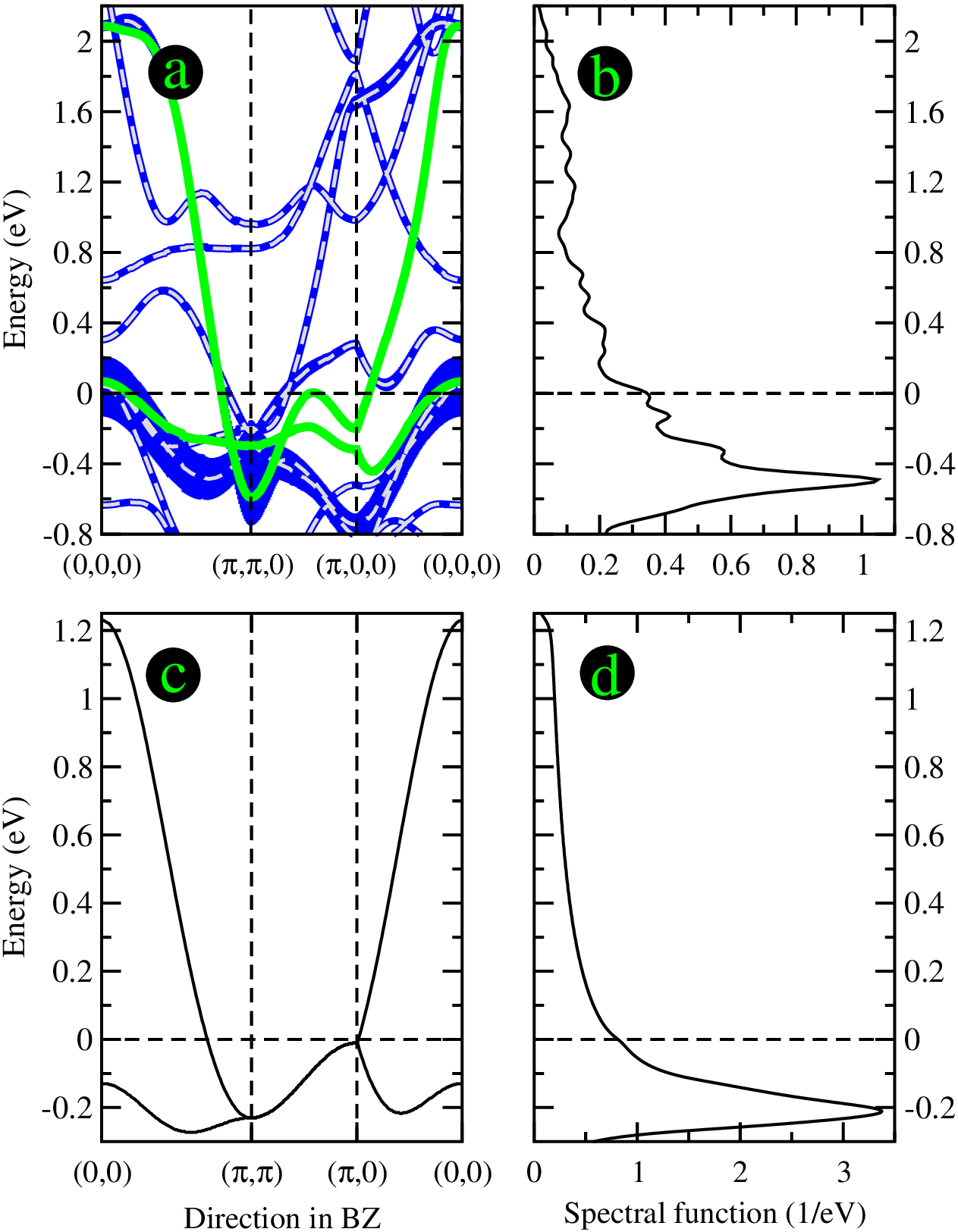}
\caption{(Color online) Band structure and spectral functions computed for BaFe$_{2}$As$_{2}$; the Fermi
energy corresponds to 0~eV. (a) Dispersion curves calculated within LDA (dashed curves), contributions of
the orbitals with $d_{xy}$ symmetry (fat bands), energy bands of a two-orbital model obtained as a projection
onto the $d_{xy}$ states (solid curves). (b) Spectral function of the $d_{xy}$ orbital from LDA. (c) Energy
bands of a model corresponding to the real hopping parameters. (d) Spectral function of the model Hamiltonian
(\ref{2DH}).}
\label{DOSBands2D}
\end{figure}

Finally we address the question concerning the microscopic origin of the peaks in the spectral function
below the Fermi level. In the following analysis we focus on the $d_{xy}$ orbital since the peak in its
spectral function is sharper than those of the other orbitals. The contributions of the $d_{xy}$ states
to the band structure are shown in the left upper panel of Fig.~\ref{DOSBands2D} as "fat bands" (i.e., the
thickness of a band is proportional to the contribution of states with selected symmetry). The peak in the
$d_{xy}$ spectral function (Fig.~\ref{DOSBands2D}, right upper panel) is formed by regions of relatively
flat bands centered at the energy approximately -0.4~eV relative to the Fermi level.

To construct a minimal model describing the energy dispersion of the $d_{xy}$ states we solve an effective
two-band Hamiltonian $H^{2D}({\bf k})$ with two Fe atoms in the unit cell, which is obtained from a projection
of Bloch states in the  vicinity of the Fermi energy onto a subspace of Wannier functions with $d_{xy}$
symmetry. The energy bands of $H^{2D}({\bf k})$ are shown in the left upper panel of Fig.~\ref{DOSBands2D}
by solid curves. In the next step we introduce the real-space Hamiltonian $H^{2D}_{real}$ written for a
square lattice with two atoms in the unit cell and nonzero hoppings within three coordinate spheres. The
Hamiltonian $H^{2D}_{real}$ has the form
\begin{equation}
H^{2D}_{real}=t\sum_{i{\mathbf R}}c^{\dagger}_{i\mathbf R}c_{i}+
t^{\prime}\sum_{i{\mathbf R}^\prime}c^{\dagger}_{i{\mathbf R}^\prime}c_{i}+
t^{\prime\prime}\sum_{i{\mathbf R}^{\prime\prime}}c^{\dagger}_{i{\mathbf R}^{\prime\prime}}c_{i},
\label{2DH}
\end{equation}
where $i$ labels the atoms, and the radius vectors ${\mathbf R}$, ${\mathbf R}^\prime$ and ${\mathbf R}^{\prime\prime}$
correspond to the cluster of nearest, next-nearest and next-next-nearest neighbors, respectively, around
atom $i$. The hopping parameters $t$=-170~meV, $t^{\prime}$=98~meV and $t^{\prime\prime}$=21~meV were
computed as Fourier transforms of the material-specific Hamiltonian $H^{2D}({\bf k})$.

The shape of the energy bands and the spectral function computed for Hamiltonian (\ref{2DH}) are shown in
the lower panels of Fig.~\ref{DOSBands2D}. They are in good agreement with the corresponding characteristics
obtained from the direct calculation. As in the multi-orbital case, the band structure of (\ref{2DH})
represents a combination of dispersive bands and bands with less pronounced ${\bf k}$-dependence.
In particular, it follows that the peak below the Fermi energy is formed by a relatively flat band located
in the same energy interval as the peak.
\begin{figure}[!h]
\centering \vspace{0.0mm}
\includegraphics[width=0.90\linewidth,angle=0]{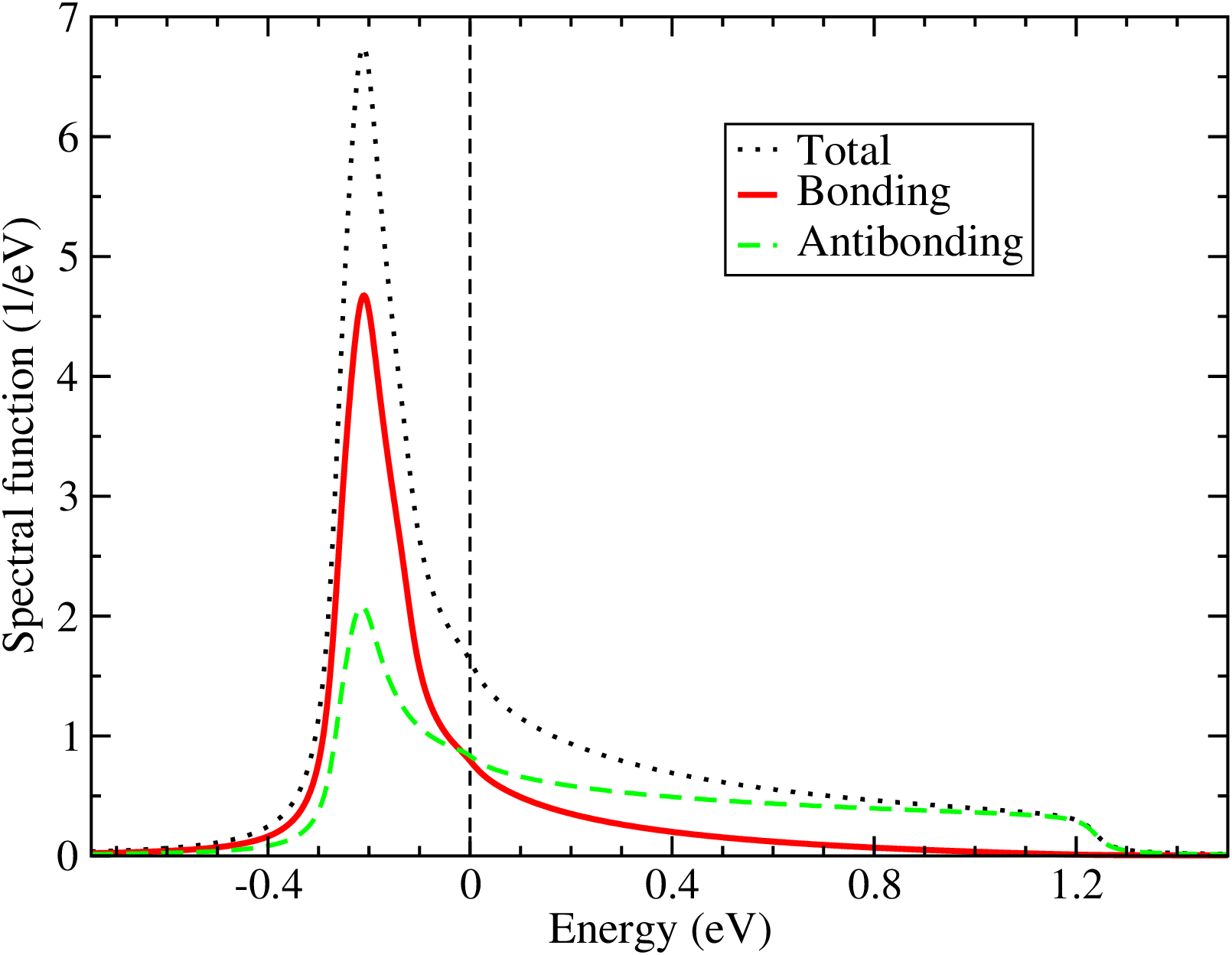}
\caption{(Color online) Spectral functions of Hamiltonian (\ref{2DH}) calculated in the basis of bonding
(solid curve) and antibonding (dashed curve) wave functions. The total spectral function is shown by a
dotted line. The Fermi energy is set to zero.}
\label{DOSbab}
\end{figure}

In order to understand the coexistence of the flat and dispersive regions within a band of given symmetry
($d_{xy}$ in our case) it is instructive to plot the spectral functions of the Hamiltonian (\ref{2DH}) in
the basis of bonding and antibonding wave functions. In other words, if $\left |\phi_{1}\right>$ is the
wave function on the one atom in the unit cell, $\left |\phi_{2}\right>$ is the wave function of the other
atom. Then the new basis is defined as
$\left |\phi_{+}\right>=\left |\phi_{1}\right>+\left |\phi_{2}\right>$ and
$\left |\phi_{-}\right>=\left |\phi_{1}\right>-\left |\phi_{2}\right>$. Spectral functions of the
Hamiltonian (\ref{2DH}) computed for that basis with realistic hopping parameters and chemical potential
are shown in Fig.~\ref{DOSbab}. From Fig.~\ref{DOSbab} it follows that the band with a pronounced dispersion
is mainly formed by antibonding linear combination $\left |\phi_{-}\right>$ and the main contribution to
the peak is provided by the bonding function $\left |\phi_{+}\right>$ whose dispersion is less pronounced.

\section{CONCLUSIONS}
By employing the LDA+DMFT method we investigated the interplay between the spectral and magnetic properties
of BaFe$_2$As$_2$. The calculated temperature dependence of the uniform magnetic susceptibility is in good
agreement with experimental data. Our calculations show that there are pronounced, temperature sensitive
peaks below the Fermi energy in the spectral function of BaFe$_2$As$_2$. We proposed a scenario according
to which the temperature increase of the susceptibility is a consequence of the thermal excitation of the
electronic states which lead to these peaks. Our analysis is based on the DMFT solution of a one-band model
with a density of states corresponding to the Fe-$d_{xy}$ spectral function of BaFe$_2$As$_2$. The results
clearly demonstrate that the peak in the spectral function in the vicinity of the Fermi energy is a prerequisite
for the linear temperature increase of magnetic susceptibility. The peaks in the real compound are due to
the weak dispersion of bonding states arising from the layered structure of BaFe$_2$As$_2$.

\section{ACKNOWLEDGMENTS}
The authors thank P.~Werner and J.~Kune\v s for supplying us with the CT-QMC code used in our calculations,
and X.H.~Chen for providing the experimental data in digitalized form. This work was supported by the Russian
Foundation for Basic Research (Projects No.~10-02-00046-a, No.~10-02-96011-r\verb"_"ural\verb"_"a,
No.~12-02-91371-CT\verb"_"a). S.L.S. is grateful to the Dynasty Foundation for support. S.L.S and V.I.A. are
grateful to the Center for Electronic Correlations and Magnetism, University of Augsburg for hospitality.
This work was supported in part by the Deutsche Forschungsgemeinschaft through Transregio TRR 80.

\end{document}